\documentstyle[aps,12pt]{revtex}

\begin{document}
\title{Phase diagram of two-species Bose-Einstein condensates in an optical lattice}
\author{G. -P. Zheng$^1$, J. -Q. Liang$^1$, and W. M. Liu$^2$}
\address{1. Institute of Theoretical Physics and Department of Physics, Shanxi\\
University, Taiyuan, Shanxi 030006, China\\
2. Joint Laboratory of Advanced Technology in Measurements, Beijing National%
\\
Laboratory for Condensed Matter Physics, Institute of Physics, Chinese\\
Academy of Sciences, Beijing 100080, China}
\date{\today}
\maketitle

\begin{abstract}
The exact macroscopic wave functions of two-species Bose-Einstein
condensates in an optical lattice beyond the tight-binding approximation are
studied by solving the coupled nonlinear Schr\"{o}dinger equations. The
phase diagram for superfluid and insulator phases of the condensates is
determined analytically according to the macroscopic wave functions of the
condensates, which are seen to be traveling matter waves.
\end{abstract}


\section{Introduction}

Since the realization of Bose-Einstein condensates (BECs) in dilute atomic
gases, a number of interesting experiments have been conducted to
investigate multispecies Bose gases, in which two or more states of
condensates exist together in a magnetic or optical trap \cite
{myatt97,stam98}. Recently, vortex states have been obtained in a
two-species Bose gas \cite{mat99}. Progress in the experiments exploring
dilute mixtures of quantum gases has stimulated intensive research on the
properties of mixed Bose gases at zero temperature \cite{ho96,esr97,pu98}
and finite temperature as well \cite{shi00}.

The BEC trapped in an optical lattice exhibites a novel feature, namely the
quantum phase transition between a Mott-insulator and a superfluid \cite
{greiner02}. Such quantum phase transition has attracted considerable
attention in recent years. As a matter of fact, atomic gas of bosons in BEC
subjected to a lattice potential which is turned on smoothly can be kept in
the superfluid phase as long as the atom-atom interactions are small
comparing with the tunnel coupling. In this regime, the kinetic energy is
dominant in the total energy of the boson system. With an increase of the
potential depth of the optical lattice, it is getting more and more
difficult for bosons to tunnel from one site to the other, and finally the
system attends an insulator phase above a critical value of the potential
depth. In this case, the phase coherence is absent and the number of boson
atoms in each lattice site becomes the same. The system possesses a
Mott-insulator behavior. Various approaches have been proposed to understand
theoretically the quantum phase transition and to determine the phase
diagram as a function of BEC parameters \cite
{fisher89,jaksch98,oosten01,chen03,jun03,xie04}.

Motivated\ by both the experimental and theoretical progress, we in the
present paper study the phase diagram\ for superfluid and insulator phases
of two-species BECs in a one-dimensional (1D) optical lattice and the
property of persistent current as well. The paper is organized as follows.
In Sec. II, the exact macroscopic wave functions of the condensates which
are not in the tight-binding regime are constructed by solving the coupled
nonlinear Schr\"{o}dinger equations. In Sec. III, the phase diagram is
determined analytically according to the macroscopic wave functions of the
condensates, i.e., the order parameters. Finally, we summarize our results
in Sec. IV.

\section{The exact macroscopic wave functions}

In this 1D geometry, the confinement along the radial direction is so tight
that the trap frequency $\omega _0$ along the radial direction is much
greater than the mean-field interaction energy. At low temperatures, the
dynamics of the atoms in the radial direction is essentially ``frozen,''
with all the atoms occupying the ground state of the harmonic trap\ with the
wave function that 
\begin{equation}
\phi _0\left( y,z\right) =\sqrt{\frac 1{\pi l_0^2}}\exp \left[ -\left(
y^2+z^2\right) /2l_0^2\right] \text{.}
\end{equation}
Here the extension of the wave function in the radial direction is given by
the length\ scale $l_0\equiv \sqrt{\hbar /m\omega _0}$ of a harmonic
oscillator, where $m$ is the mass of the atoms.

In the mean-field regime, $l_0$ is much greater than the radius of the
interatomic potential $R_e$. The scattering of atoms in this effective 1D
system is thus still a process of three-dimension. According to ref. \cite
{olsh98}, the effective coupling constant in this 1D system is 
\begin{equation}
g_{1D}=\frac{2\hbar ^2}m\frac a{l_0\left( l_0-Ca\right) }\text{,}
\end{equation}
where $a$ is the {\it s-}wave scattering length and $C$ is a numerical
constant of the order unit. The term $Ca$ in Eq. (2) is negligible for $%
l_0>>R_e$. In this limit, the expression for $g_{1D}$ is the same as that
obtained by averaging over the radial wave function (1),

\begin{equation}
g_{1D}=g_{3D}\int_{-\infty }^\infty \int_{-\infty }^\infty dydz\phi
_0^4\left( y,z\right) =\frac{2\hbar ^2a}{ml_0^2}\text{.}
\end{equation}
We use this expression in the rest of this paper.

We consider the two-species BECs in a 1D periodic potential. The energy
functional is seen to be 
\begin{eqnarray}
E[\psi _1,\psi _2] &=&\int dx\left\{ \sum_{i=1,2}\left[ \frac{\hbar ^2}{2m_i}%
\left| \frac{\partial \psi _i}{\partial x}\right| ^2+V_i(x)|\psi
_i|^2\right. \right.   \nonumber \\
&&\left. \left. +\frac{\hbar ^2a_i}{m_il_i^2}|\psi _i|^4\right] +\frac{%
2\hbar ^2a_{12}}{\sqrt{m_1m_2}l_1l_2}|\psi _1|^2|\psi _2|^2\right\} \text{,}
\end{eqnarray}
where $\psi _i$, $m_i$, and $l_i=\sqrt{\hbar /m_i\omega _0}$ are the
macroscopic wave functions of the condensates, the mass, and the
harmonic-oscillator lengths in the radial direction of the $i$th species $%
\left( i=1,2\right) $, respectively. $a_1$, $a_2$, and $a_{12}$ denote the 
{\it s-}wave scattering lengths between same-species and interspecies
collisions. $V_i(x)$ are the periodic potentials, 
\begin{equation}
V_i(x)=V_{0,i}\text{sn}^2\left( k_Lx,k\right) \text{,}
\end{equation}
with $V_{0,i}$ denoting the magnitude of potentials, where $k_L=2\pi
/\lambda $ is the wavevector of the laser light and $\lambda $ is the
wavelength, corresponding to a lattice period $d=\lambda /2$. sn$(k_Lx,k)$
is the Jacobian elliptic sine function with modulus $k$ $\left( 0\leq k\leq
1\right) $. In the limit $k=0$, the Jacobian\ elliptic sine reduces to the
sinusoid function and thus $V(x)$ possesses a standard form of the standing
light wave. For values of $k<0.9$, the potential is virtually
indistinguishable from a standing light wave. Finally, for $k\rightarrow 1$, 
$V(x)$ becomes an array of well-separated hyperbolic secant potential
barriers or wells.

The governing equations of the trapped BECs are obtained in terms of the
variational procedure \cite{dalfovo99},

\begin{equation}
i\hbar \frac{\partial \psi _i}{\partial t}=\frac{\delta E}{\delta \psi _i^{*}%
}\text{,}
\end{equation}
which leads to the coupled nonlinear Schr\"{o}dinger equations 
\begin{eqnarray}
i\hbar \frac{\partial \psi _1}{\partial t} &=&-\frac{\hbar ^2}{2m_1}\frac{%
\partial ^2\psi _1}{\partial x^2}+\frac{2\hbar ^2a_1}{m_1l_1^2}|\psi
_1|^2\psi _1  \nonumber \\
&&+\frac{2\hbar ^2a_{12}}{\sqrt{m_1m_2}l_1l_2}|\psi _2|^2\psi _1+V_1(x)\psi
_1\text{,}  \nonumber \\
i\hbar \frac{\partial \psi _2}{\partial t} &=&-\frac{\hbar ^2}{2m_2}\frac{%
\partial ^2\psi _2}{\partial x^2}+\frac{2\hbar ^2a_{12}}{\sqrt{m_1m_2}l_1l_2}%
|\psi _1|^2\psi _2  \nonumber \\
&&+\frac{2\hbar ^2a_2}{m_2l_2^2}|\psi _2|^2\psi _2+V_2(x)\psi _2\text{.}
\end{eqnarray}

For the case of weakly coupled condensates in an optical lattice \cite
{cata01}, the wave function $\psi $ can be decomposed as a sum of wave
functions localized in each well of the periodic potential (tight-binding
approximation) with the assumption relying on the fact that the height of
the interwell barrier is much higher than the chemical potential. We,
however, do not restrict ourselves to the low-energy case and look for the
global condensate wave functions of excitations: $\psi _i(x,t)=\phi
_i(x)\exp \left( -i\mu _it/\hbar \right) $, where $\mu _i$ $(i=1,2)$ are the
chemical potentials. Thus the spatial wave functions satisfy the stationary
coupled nonlinear Schr\"{o}dinger equations that

\begin{eqnarray}
\mu _1\phi _1 &=&-\frac{\hbar ^2}{2m_1}\frac{\partial ^2\phi _1}{\partial x^2%
}+\frac{2\hbar ^2a_1}{m_1l_1^2}|\phi _1|^2\phi _1  \label{sf} \\
&&+\frac{2\hbar ^2a_{12}}{\sqrt{m_1m_2}l_1l_2}|\phi _2|^2\phi _1+V_1(x)\phi
_1\text{,}  \nonumber \\
\mu _2\phi _2 &=&-\frac{\hbar ^2}{2m_2}\frac{\partial ^2\phi _2}{\partial x^2%
}+\frac{2\hbar ^2a_{12}}{\sqrt{m_1m_2}l_1l_2}|\phi _1|^2\phi _2  \nonumber \\
&&+\frac{2\hbar ^2a_2}{m_2l_2^2}|\phi _2|^2\phi _2+V_2(x)\phi _2\text{.} 
\nonumber
\end{eqnarray}

With the general form of spatial wave functions $\phi _i(x)$ written as \cite
{carr00} $\phi _i(x)=r_i(x)\exp \left[ i\varphi _i\left( x\right) \right] $,
Eq. (\ref{sf})\ can be separated as real and imaginary parts. We then
integrate once for the imaginary part and obtain the first-order
differential equations for the phases $\varphi _i(x)$, 
\begin{equation}
\varphi _i^{^{\prime }}(x)=\frac{\alpha _i}{r_i^2(x)}\text{,}  \label{pf}
\end{equation}
where parameters $\alpha _i$ $(i=1,2)$ are constants of integration to be
determined. Substituting Eq. (\ref{pf}) into the real part obtained from Eq.
(\ref{sf}) and integrating again, we find

\begin{eqnarray}
\left( r_1r_1^{^{\prime }}\right) ^2 &=&\frac{2a_1}{l_1^2}r_1^6-\frac{%
2m_1\mu _1}{\hbar ^2}r_1^4+\beta _1r_1^2-\alpha _1^2  \label{rp} \\
&&+\frac{4a_{12}\sqrt{m_1}}{\sqrt{m_2}l_1l_2}r_1^2\int r_2^2d\left(
r_1^2\right)  \nonumber \\
&&+\frac{2m_1}{\hbar ^2}r_1^2\int V_1\left( x\right) d\left( r_1^2\right) 
\text{,}  \nonumber \\
\left( r_2r_2^{^{\prime }}\right) ^2 &=&\frac{2a_2}{l_2^2}r_2^6-\frac{%
2m_2\mu _2}{\hbar ^2}r_2^4+\beta _2r_2^2-\alpha _2^2  \nonumber \\
&&+\frac{4a_{12}\sqrt{m_2}}{\sqrt{m_1}l_1l_2}r_2^2\int r_1^2d\left(
r_2^2\right)  \nonumber \\
&&+\frac{2m_2}{\hbar ^2}r_2^2\int V_2\left( x\right) d\left( r_2^2\right) 
\text{,}  \nonumber
\end{eqnarray}
where $\beta _i$ $(i=1,2)$ denote additional constants of integration.

We then construct the solutions as 
\begin{equation}
r_i^2(x)=A_i\text{sn}^2\left( k_Lx,k\right) +B_i\text{,}  \label{cs}
\end{equation}
where the constants $B_i$ $(i=1,2)$ determine the mean amplitudes and act as
the dc offsets for the numbers of the condensed atoms \cite{bronski01}, and
the parameters $A_i$ $(i=1,2)$ are to be determined.

Substituting Eq. (\ref{cs}) into Eq. (\ref{rp}) and using identities of
Jacobian elliptic functions, we obtain eight equations for the parameters $%
\alpha _i$, $\beta _i$, $\mu _i$, and $A_i$. Eliminating $\beta _i$, we find 
\begin{eqnarray}
A_1 &=&\frac{\frac{\sqrt{m_1}l_1l_2a_{12}}{\sqrt{m_2}}\left(
m_2V_{0,2}-\hbar ^2k_L^2k^2\right) -a_2l_1^2\left( m_1V_{0,1}-\hbar
^2k_L^2k^2\right) }{2\hbar ^2\left( a_1a_2-a_{12}^2\right) }\text{,} 
\nonumber \\
A_2 &=&\frac{\frac{\sqrt{m_2}l_1l_2a_{12}}{\sqrt{m_1}}\left(
m_1V_{0,1}-\hbar ^2k_L^2k^2\right) -a_1l_2^2\left( m_2V_{0,2}-\hbar
^2k_L^2k^2\right) }{2\hbar ^2\left( a_1a_2-a_{12}^2\right) }\text{,} 
\nonumber \\
&&~~
\end{eqnarray}
and 
\begin{eqnarray}
\alpha _1^2 &=&B_1k_L^2\left[ \frac{k^2}{A_1}B_1^2\!+\!\left( 1+k^2\right)
B_1\!+\!A_1\right] \text{,} \\
\alpha _2^2 &=&B_2k_L^2\left[ \frac{k^2}{A_2}B_2^2\!+\!\left( 1+k^2\right)
B_2\!+\!A_2\right] \text{,}  \nonumber
\end{eqnarray}
and 
\begin{eqnarray}
\mu _1 &=&\frac{\hbar ^2k_L^2}{2m_1}\left( 1+k^2+\frac{6a_1}{l_1^2k_L^2}B_1+%
\frac{4a_{12}\sqrt{m_1}}{l_1l_2k_L^2\sqrt{m_2}}B_2\right.   \nonumber \\
&&\left. +\frac{2a_{12}\sqrt{m_1}}{l_1l_2k_L^2\sqrt{m_2}}\frac{A_2}{A_1}B_1+%
\frac{m_1V_{0,1}}{\hbar ^2k_L^2}\frac{B_1}{A_1}\right) \text{,}  \nonumber \\
\mu _2 &=&\frac{\hbar ^2k_L^2}{2m_2}\left( 1+k^2+\frac{6a_2}{l_2^2k_L^2}B_2+%
\frac{4a_{12}\sqrt{m_2}}{l_1l_2k_L^2\sqrt{m_1}}B_1\right.   \nonumber \\
&&\left. +\frac{2a_{12}\sqrt{m_2}}{l_1l_2k_L^2\sqrt{m_1}}\frac{A_1}{A_2}B_2+%
\frac{m_2V_{0,2}}{\hbar ^2k_L^2}\frac{B_2}{A_2}\right) \text{.}
\end{eqnarray}

For $k=0$, $\text{sn}(k_Lx,0)=\sin (k_Lx)$, the solutions reduce to 
\begin{equation}
\psi _i(x,t)=\sqrt{A_i^0\sin ^2\left( k_Lx\right) +B_i}\exp \left\{ i\left[
\varphi _i^0\left( x\right) -\mu _i^0t/\hbar \right] \right\} \text{,}
\label{wf}
\end{equation}
where 
\begin{eqnarray}
A_1^0 &=&\frac{\sqrt{m_1m_2}a_{12}l_1l_2V_{0,2}-m_1a_2V_{0,1}l_1^2}{2\hbar
^2\left( a_1a_2-a_{12}^2\right) }, \\
A_2^0 &=&\frac{\sqrt{m_2m_1}a_{12}l_1l_2V_{0,1}-m_2a_1V_{0,2}l_2^2}{2\hbar
^2\left( a_1a_2-a_{12}^2\right) }.  \nonumber
\end{eqnarray}

The phases $\varphi _i^0\left( x\right) $ $(i=1,2)$ are determined by
nonlinear equations 
\begin{equation}
\tan \left[ \varphi _i^0\left( x\right) \right] =\pm \sqrt{1+\frac{A_i^0}{B_i%
}}\tan \left( k_Lx\right) 
\end{equation}
and 
\begin{eqnarray}
\mu _1^0 &=&\frac{\hbar ^2k_L^2}{2m_1}\left( 1+\frac{6a_1}{l_1^2k_L^2}B_1+%
\frac{4a_{12}\sqrt{m_1}}{l_1l_2k_L^2\sqrt{m_2}}B_2\right.   \nonumber \\
&&\left. +\frac{2a_{12}\sqrt{m_1}}{l_1l_2k_L^2\sqrt{m_2}}\frac{A_2^0}{A_1^0}%
B_1+\frac{m_1V_{0,1}}{\hbar ^2k_L^2}\frac{B_1}{A_1^0}\right) \text{,} 
\nonumber \\
\mu _2^0 &=&\frac{\hbar ^2k_L^2}{2m_2}\left( 1+\frac{6a_2}{l_2^2k_L^2}B_2+%
\frac{4a_{12}\sqrt{m_2}}{l_1l_2k_L^2\sqrt{m_1}}B_1\right.   \nonumber \\
&&\left. +\frac{2a_{12}\sqrt{m_2}}{l_1l_2k_L^2\sqrt{m_1}}\frac{A_1^0}{A_2^0}%
B_2+\frac{m_2V_{0,2}}{\hbar ^2k_L^2}\frac{B_2}{A_2^0}\right) \text{.}
\end{eqnarray}
The constants $A$ and $B$ are related by restrictions such that $B_1\geq
-A_1^0$ for $A_1^0<0$ and $B_1\geq 0$ for $A_1^0>0$; $B_2\geq -A_2^0$ for $%
A_2^0<0$ and $B_2\geq 0$ for $A_2^0>0$.

\section{The phase diagram}

The average particle number densities $n_i$ for the two species are obtained
as 
\begin{eqnarray}
n_i &=&\frac 1L\int_0^L\left| \psi _i\left( x,t\right) \right| ^2dx \\
&=&\frac 1{h\pi }\int_0^{h\pi }\left[ A_i^0\sin ^2\left( x^{^{\prime
}}\right) +B_i\right] dx^{^{\prime }}\text{,}  \nonumber
\end{eqnarray}
where $x^{^{\prime }}=k_Lx$ and $L=hd$ denotes the length of the optical
lattice with $h=1,2,3,...$ This leads to 
\begin{equation}
B_i=n_i-\frac{A_i^0}2\text{.}
\end{equation}

Then the macroscopic wave functions of the condensates Eq. (\ref{wf}) can
exist only when

\begin{eqnarray}
n_1 &\geq &\frac{|A_1^0|}2=\frac{\left| a_{12}V_{0,2}-a_2V_{0,1}\right| }{%
4\hbar \omega _0\left| a_1a_2-a_{12}^2\right| }\text{,} \\
n_2 &\geq &\frac{|A_2^0|}2=\frac{\left| a_{12}V_{0,1}-a_1V_{0,2}\right| }{%
4\hbar \omega _0\left| a_1a_2-a_{12}^2\right| }\text{.}  \nonumber
\end{eqnarray}

The condensate atom currents can be evaluated from the usual definition, $%
j=(\hbar /m)$Im$\left[ \psi ^{*}\left( \partial \psi /\partial x\right)
\right] ${\bf \ }\cite{choi99}{\bf , }with the exact wave functions Eq. (\ref
{wf}) which are seen to be travelling matter waves. The result is 
\begin{eqnarray}
j_i &=&\pm \frac{\hbar k_L}{m_i}\sqrt{B_i(B_i+A_i^0)}  \label{pc} \\
&=&\pm \frac{\hbar k_L}{m_i}\sqrt{n_i^2-\frac{(A_i^0)^2}4}\text{,}  \nonumber
\end{eqnarray}
which are independent of space-time variables and therefore persistent
currents. We may demand that the wave functions Eq. (\ref{wf}) satisfy the
periodic boundary condition{\bf \ }$\psi _i(x,t)=\psi _i(x+L,t)$\ which is
naturally fulfilled as the total length $L$\ is an integer times the lattice
constant $d${\bf \ (}$L=hd,h=1,2,3,...${\bf ). }These periodic solutions in
1D space with spatial period $L$\ are equivalent to the solutions in a ring
of circumstance $L$. The persistent currents then can be viewed as in the
optical lattice ring.

It is found from Eq. (\ref{pc}) that the persistent currents are valid only
for conditions in which the number density of atoms is greater than critical
values, $n_i\geq |A_i^0|/2$, i.e., when the macroscopic wave functions of
the condensates exist. These persistent currents are similar to the 1D Fr%
\"{o}hlich superconductivity induced by the traveling lattice wave \cite
{frohlich54} and can be controlled by adjusting the barriers height of the
periodic potentials and parameters of the bosonic atoms. The currents
increase with the decrease of $|A_i^0|$\ and approach the asymptotic maximum
values $j_{i,\max }=\pm \hbar n_ik_L/m_i$\ when $|A_i^0|$\ become
vanishingly small. With the recent progress made on confinement of atoms in
the light-induced \cite{olsh93,renn95} as well as the magnetic-field-induced 
\cite{dens99,lean02} atom waveguides \cite{kase02}, the persistent currents
may be observed experimentally in the future.

The energy spectrum for the two species is obtained as 
\begin{eqnarray}
\mu _1^0 &=&E_{R,1}+\frac{V_{0,1}}2+2\hbar \omega _0(a_1n_1+a_{12}n_2)\text{,%
}  \nonumber \\
\mu _2^0 &=&E_{R,2}+\frac{V_{0,2}}2+2\hbar \omega _0(a_2n_2+a_{12}n_1)\text{,%
}
\end{eqnarray}
where $E_{R,i}=\hbar ^2k_L^2/2m_i$ are the recoil energy of an atom
absorbing one of the lattice phonons \cite{cata01}.

In this paper, we consider the two species both with repulsive interactions,
namely, $a_1>0$, $a_2>0$, $a_{12}>0$. Then the macroscopic wave functions of
the condensates can exist only when 
\begin{eqnarray}
\mu _1^0\! &\geq &E_{R,1}+\frac{V_{0,1}}2+\frac{a_1\left|
a_{12}V_{0,2}-a_2V_{0,1}\right| }{2\left| a_1a_2-a_{12}^2\right| }
\label{cp} \\
&&+\frac{a_{12}\left| a_{12}V_{0,1}-a_1V_{0,2}\right| }{2\left|
a_1a_2-a_{12}^2\right| }\text{,}  \nonumber \\
\mu _2^0\! &\geq &E_{R,2}+\frac{V_{0,2}}2+\frac{a_2\left|
a_{12}V_{0,1}-a_1V_{0,2}\right| }{2\left| a_1a_2-a_{12}^2\right| }  \nonumber
\\
&&+\frac{a_{12}\left| a_{12}V_{0,2}-a_2V_{0,1}\right| }{2\left|
a_1a_2-a_{12}^2\right| }\text{.}  \nonumber
\end{eqnarray}

The macroscopic wave functions of the condensates $\psi _i(x,t)$ are complex
functions defined as the expectation value of the boson field operators: $%
\psi _i(x,t)\equiv \left\langle \hat{\Psi}_i(x,t)\right\rangle $, which have
the meaning of order parameters and characterize the off-diagonal long-range
behavior of the one-particle density matrix $\rho _i(x^{^{\prime
}},x,t)=\left\langle \hat{\Psi}_i^{+}(x^{^{\prime }},t)\hat{\Psi}%
_i(x,t)\right\rangle $ \cite{dalfovo99}. So the condensates can be kept in
the superflud phase only when the macroscopic wave functions exist.
Otherwise, the phase coherence and the currents vanish and therefore the
condensates are in the insulator phase. Then we obtain four cases of phases
for two-species BECs in a 1D optical lattice as follows.

{\it Case 1}{\bf .} The two species are both in the superfluid phase, namely

\begin{eqnarray}
n_1 &\geq &\frac{\left| a_{12}V_{0,2}-a_2V_{0,1}\right| }{4\hbar \omega
_0\left| a_1a_2-a_{12}^2\right| }\text{,} \\
n_2 &\geq &\frac{\left| a_{12}V_{0,1}-a_1V_{0,2}\right| }{4\hbar \omega
_0\left| a_1a_2-a_{12}^2\right| }\text{,}  \nonumber
\end{eqnarray}
which is labeled as SS in the phase diagram, Fig. 1.

{\it Case 2}{\bf .} Species 1 is in the superfluid phase while species 2 is
in the insulator phase,

\begin{eqnarray}
n_1 &\geq &\frac{\left| a_{12}V_{0,2}-a_2V_{0,1}\right| }{4\hbar \omega
_0\left| a_1a_2-a_{12}^2\right| }\text{,} \\
n_2 &\leq &\frac{\left| a_{12}V_{0,1}-a_1V_{0,2}\right| }{4\hbar \omega
_0\left| a_1a_2-a_{12}^2\right| }\text{,}  \nonumber
\end{eqnarray}
labeled as SI.

{\it Case 3}{\bf .} Species 1 is in the insulator phase and species 2 is in
the superfluid phase,

\begin{eqnarray}
n_1 &\leq &\frac{\left| a_{12}V_{0,2}-a_2V_{0,1}\right| }{4\hbar \omega
_0\left| a_1a_2-a_{12}^2\right| }\text{,} \\
n_2 &\geq &\frac{\left| a_{12}V_{0,1}-a_1V_{0,2}\right| }{4\hbar \omega
_0\left| a_1a_2-a_{12}^2\right| }\text{,}  \nonumber
\end{eqnarray}
labeled as IS.

{\it Case 4}{\bf .} The two\ species are both in the insulator phase,

\begin{eqnarray}
n_1 &\leq &\frac{\left| a_{12}V_{0,2}-a_2V_{0,1}\right| }{4\hbar \omega
_0\left| a_1a_2-a_{12}^2\right| }\text{,} \\
n_2 &\leq &\frac{\left| a_{12}V_{0,1}-a_1V_{0,2}\right| }{4\hbar \omega
_0\left| a_1a_2-a_{12}^2\right| }\text{,}  \nonumber
\end{eqnarray}
labeled as II.

The\ quantum phases of the condensates can be determined by all parameters
of BECs and optical lattice as shown above. In Fig. 1, we show the phase
diagram with the various same-species {\it s}-wave scattering lengths and
equal particle number density $n_1=n_2=n$ and the magnitudes of potentials ($%
V_{0,1}=V_{0,2}=V_0$\ ) for simplicity. Thus the conditions for the two
species in the superfluid phase are given by

\begin{equation}
\frac{V_0}{4n\hbar \omega _0}\leq \frac{\left| a_1a_2-a_{12}^2\right| }{%
\left| a_{12}-a_2\right| }  \label{c1}
\end{equation}
and

\begin{equation}
\frac{V_0}{4n\hbar \omega _0}\leq \frac{\left| a_1a_2-a_{12}^2\right| }{%
\left| a_{12}-a_1\right| }\text{,}  \label{c2}
\end{equation}
respectively.

Component separation in two-species BECs\ has been predicted by means of
mean-field theory \cite{ho96,esr97,pu98,shi00} and observed in experiments 
\cite{myatt97,stam98} when the relation of the scattering lengths that $%
a_{12}>\sqrt{a_1a_2}$ is fulfilled. From the above conditions Eqs. (\ref{c1}%
), (\ref{c2}) and the phase diagram Fig. 1, we find that the larger values
of $a_{12}$\ favor the superfluid phase in the two-species mixture and the
component separation according to the experimental observation\cite
{myatt97,stam98}. Particularly when the interspecies scattering length
approaches the value of the same-species such that $a_{12}=a_2$\ or $a_1$,\
the conditions Eqs. (\ref{c1}), \ref{c2}) result in the superfluid phase
independent of the potential magnitude $V_0$. One should not be surprised by
this result since we consider the case in which the chemical potential is
always higher than the potential magnitude $V_0$\ seen from Eq. (\ref{cp}).

\section{Conclusion}

In conclusion, the exact macroscopic wave functions of two-species BECs in
an optical lattice beyond the tight-binding approximation are studied. The
phase diagram is determined analytically according to the order parameters,
and persistent currents in an optical lattice ring are obtained explicitly
in terms of the exact wave functions, which are seen to be traveling matter
waves.

\section{Acknowledgments}

This work was supported by the NSF of China under Grants No. 10475053, No.
60490280, No. 90406017, and No. 90403034.

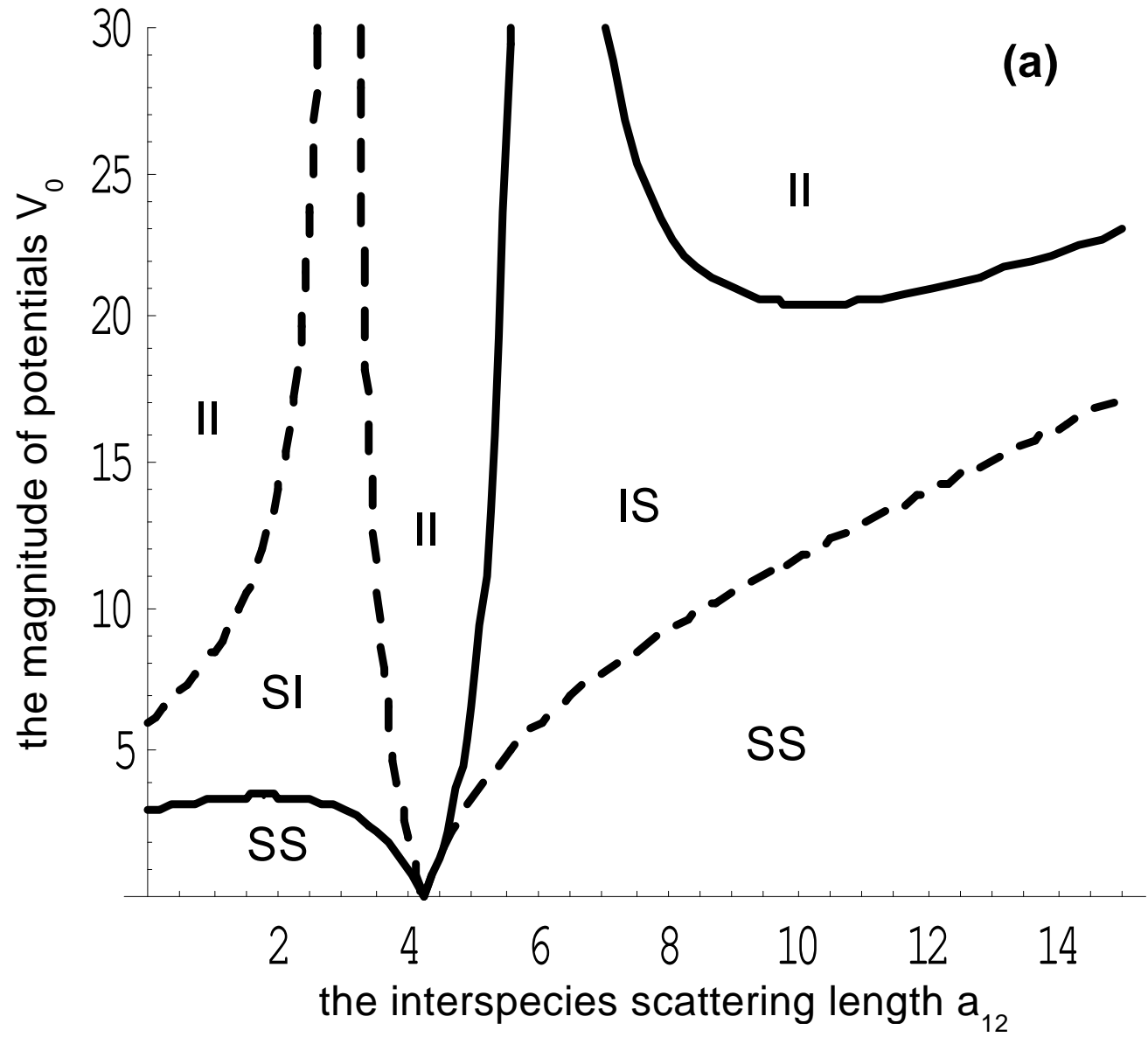
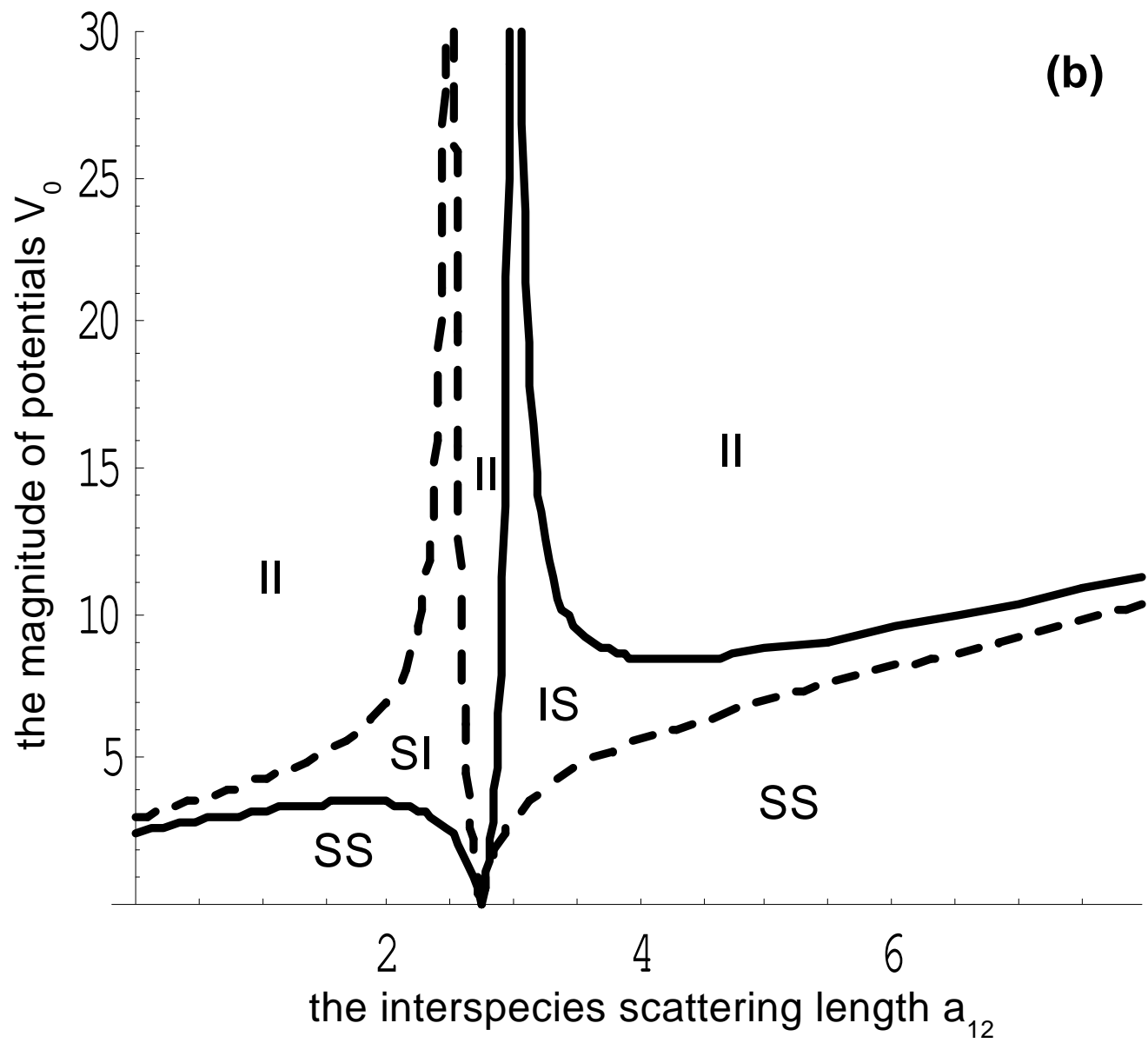
\end{document}